# Exploring electrical conductivity within mesoscopic phases of semiconducting PEDOT:PSS films by Broadband Dielectric Spectroscopy

**A. N.Papathanassiou [a] (*), I. Sakellis [a, b], J. Grammatikakis [a], E. Vitoratos [c] and S. Sakkopoulos [c]**

[a] University of Athens, Physics Department, Solid State Physics Section, Panepistimiopolis, GR15784 Zografos, Athens, Greece

[b] National Center for Scientific Research 'Demokritos', Institute for Advanced Materials, Physicochemical Processes, Nanotechnology & Microsystems, , Athens, Greece

[c] University of Patras, Physics Department, Condensed Matter Physics Section, Patras, Greece

**Abstract**

Poly(3,4-ethylenedioxythiophene) : poly(styrene sulfonic acid (PEDOT:PSS), an optically transparent organic semi-conductor, constitutes a suspension of conducting PEDOT:PSS grains, shelled by an insulating layer of PSS. While a percolation network enhances dc conductivity, structural and electrical inhomogenity hinters electric charge flow giving rise to capacitance effects. In the present work, two distinct relaxation mechanisms are detected by Broadband Dielectric Spectroscopy (BDS). Double polarization mechamisms are predicted by bi-phase suspension dielectric theory. Within the frame of interfacial polarization, we propose a methodology to have an insight into the local conductivity of the interior of mesoscopic conducting phase.

**PACS indexes:** 77.22.Ch; 77.22.Gm; 73.61.Ph

Poly(3,4-ethylenedioxythiophene) (PEDOT) is a p-type conjugated polymer which is optically transparent in the visible region and can become good electrical conductor by proper synthesis or processing [1]. PEDOT exhibits low solubility by itself, but, when water polyelectrolyte (PSS) solution is used as a charge balancing dopant, PEDOT:PSS mixture form in water a micro-dispersion that forms thin transparent films by the spin-coating technique. Spin coated films of PEDOT:PSS is a phase-segregated material consisting of lentil-like PEDOT:PSS grains of about 25 nm width, which are covered by a thin layer of a few nanometers of PSS [2, 3, 4]. The conductivity in the interior of these grains is high but the total (global) conductivity of the film is reduced, since the insulating PSS shells impede electric charge low.

Structural disorder and conductivity heterogeneity are, in principle, likely to induce capacitance effect due to charge trapping. Dc conductivity measurements integrate contributions of different phases yielding an effective (volume) conductivity value and trace percolation pathways [5]. Alternatively, complex perimittivity measurements in the frequency domain, different spatio-temporal scales are explored [6, 7] and the entire energy profile can be monitored. To the best of our knowledge, complex permittivity has been used to obtain the ambient temperature dc resistivity of PEDOT:PSS as a function of the weight ratio of its two components [2] and to trace at room temperature the interfacial polarization between conductive-rich PEDOT domains and poorly conductive PSS at PEDOT:PSS electrodes sandwiching a polymer-dispersed liquid crystal layer [3]. In the present work, Broadband Dielectric Spectroscopy was used to probe the response of PEDOT:PSS films on pristine plastics (PET) substrate, probed on the open surface of it by a specially designed electrode configuration. Long range (related with the percolation network that gives rise to the dc conductivity) and localized electric charge flow within different mesoscopic phases. Furthermore, the activation energy for the conductivity *inside* these phases is obtained by studying the temperature variation of the relaxation spectra, within theoretical predictions for bi-phase suspensions.

PEDOT:PSS films with thickness 50 nm, coated on PET were made from an aqueous dispersion of PEDOT:PSS (CLEVIOS PH 500, H.C. Starck), where the ratio PEDOT-to-PSS was 1:2.5 by weight [8]. A sample holder was constructed [9] in order get the frequency response of PEDOT:PSS from its open surface: PEDOT:PSS/PET were placed on an optically flat and electrically grounded platinum surface, in a manner that PET lies on the grounded metal, while PEDOT:PSS remains an open surface. A couple of parallel silver-epoxy strips (the separation distance was 5 mm) were attached on an optically flat surface of a thick glass disk, which was spring loaded to the open surface of the PEDOT:PSS/PET. This structure allows measurement of the surface ac electrical conductivity on the open surface of (conducting or semi-conducting) films deposited on an insulating substrate (such as PET). The substrate (PET) is a good insulator (compared with the conductivity of PEDOT:PSS) and attaches a grounded metal to eliminate any spurious signal from the measuring circuit. The setup was thoroughly inspected and calibrated by performing surface ac conductivity measurements (in the frequency range from 1 mHz to 1 MHz), with the aforementioned electrode arrangement, as well as, with circular concentric electrodes, with electrodes of different materials, and bulk conductivity measurements in capacitor-type electrode sample holder, on test materials. It was found that (e.g., for a test material such as doped Si crystal) the surface ac conductance was proportional to the bulk conductivity for the entire frequency range. All conductance data presented in the present work are those recorded from the two parallel strip electrode setup and are labeled as *surface conductance*. One can correlate surface conductance with the bulk one provided that relaxation close to the surface is identical to the relaxation occurring in the bulk.

A vacuum cryostat [10], operating from the liquid nitrogen temperature to 420K, accommodated the sample holder described above. Isothermal conditions could be achieved with an accuracy of 0.2 K. A schematic diagram of a vertical cross-section of the cylindrical cryostat is depicted in Figure 1, together with a sketch of the sample holder and the strip-electrodes configuration. The cryostat ensures dark conductivity conditions. A Solartron 1240 Gain Phase Frequency Response Analyzer with a BDC pre-amplifier was used to measure complex impedance in the frequency range from 1 mHz to 1 MHz. The instruments were fully controlled by a computer using WinDeta (Novocontrol) software. The amplitude of the applied ac voltage was 50mV, Various amplitudes up to 1.5 V were also applied and linear response was found.

The imaginary part of the (relative) complex permittivity $\varepsilon^*$ is connected to complex electrical conductivity $\sigma^*$ through the relation $\text{Im}(\varepsilon^*) = \text{Re}(\sigma^*)/(2\pi f \varepsilon_0) \propto G(f)/f$, where f denotes the frequency of the harmonic voltage applied to the specimen, $\varepsilon_0$ is the permittivity of free space and G(f) is the measured conductance. A frequency independent conductance G provides a straight line with slope –1 when $G/\varepsilon_0 f$ is plotted against f, in log-log diagram. In Figure 2, $G/\varepsilon_0 f$ is plotted as a function of frequency at various temperatures. The dc contribution superimposes with dielectric loss peaks (relaxation peaks or dispersions. As can be seen in Figure 2, at room temperature, the dc component is so strong that masks dielectric relaxation peaks and all data points lie on a straight line with slope –1. By decreasing the temperature, the dc contribution is suppressed and a broad dispersion is revealed in the high frequency region. This dispersion shifts gradually toward lower frequencies on reducing the temperature.

However, further analyses of the experimental data evidence for the presence of another (low frequency) dispersion: In Figure 3, a typical isotherm is shown together with a fitting function consisting of a dc conductivity component and a couple of Debye peaks (the use of a single dispersion could not provide a good match of the fitting function to the data points in the intermediate frequency region). The reduced residuals defined as $R \equiv (y_m - y_f)/y_m$, where, for given x, $y_m$ and $y_f$ are the measured and best-fitted theoretical y-values, respectively) as a function of frequency are also presented (inset of Figure3). A dominant high frequency (HF) one has its maximum around $10^4$ Hz and another low frequency (LF) one around $10^2$ Hz. The HF peak can be correlated with that reported in [3] with maximum around $10^5$ Hz. Despite the reduction of the dc component by cooling, LF relaxation is masked significantly by the dc component and the strong neighboring HF peak. The lack of a direct observation of a clear maximum of the LF relaxation, urged us to examine whether two distinct Debye peaks or a single asymmetric relaxation contribute to the spectra. The first scenario is valid for two reasons:
(i) Different fitting functions for a unique relaxation peak (combined with the dc conductivity component) were unsuccessful to fit the crossover region from the low frequency dc regime to the high frequency dispersive one. Instead, a couple of Debye peaks could better fit the whole spectra.
(ii) The specimen was thermally aged to reduce furthermore the dc conductivity. Although the thermally induced structural modification of the polymer chains might affect the dielectric relaxation, in the spectra of an annealed specimen, we can see in Figure 4, clearly that, apart from the HF peak, LF appears clearly as a 'knee' around $10^2$ Hz.

As described in the introduction, the PEDOT:PSS film consists of lentil-like PEDOT:PSS grains of about 25 nm width, shelled by a thin layer of PSS [2, 3]. While the conductivity of interior of these grains is high, the total conductivity of the film is reduced, because the insulating PSS shells impede electric charge flow. The system can be simulated to a first approximation as a suspension of shelled conducting suspension, with conductivities $\sigma_{sp}$, $\sigma_{sh}$ and $\sigma_1$, are the conductivities of the sphere, shell and the host material, respectively, and $\sigma_{sh} \prec\prec \sigma_1 \prec \sigma_{sp}$. We consider inter-grain phase (phase 1) consisting of dispersed isolated PSS and PEDOT chain, which are not organized to a conductive phase (as they do in the interior of the grains) more conductive than the compact PSS shell, due to the isolated PEDOT chains. The model can by more refined by regarding prolate or oblate spheroids or ellipsoids.

Theoretical models and experimental results [11, 12] have established that suspensions of conducting spheres, covered by a thin layer of insulator dispersed into a less conducting medium (compared with the interior conductivity), compared with that of the inner matter, yield two distinct relaxation mechanisms in the frequency domain:
(i) A high frequency (HF) mechanism, which corresponds to the interfacial polarization of the thin-shelled conductive spheres inside the (less conductive) host matrix
(ii) A low frequency (LF) one, which corresponds to a suspension of homogeneous spheres of radius equal to the external radius of the shelled spheres and effective permittivity and conductivity equal to the corresponding effective ones of the bi-phase shelled particles.

The relaxation spectra of PEDOT:PSS films is compatible with the predictions of a three-phase suspension of shelled spheres described above. It is worth noticing that the observed maximum frequencies are comparable to with the location of the peaks observed in suspension at high volume concentration of polystyrene microcapsules filled with conducting fluids (the low frequency appears roughly in the vicinity of 10 to $10^3$ kHz and the high frequency one from 1 to 10 MHz) [11]. The frequencies where these dispersions exhibit their maximum and the intensities are quite complex functions of the bulk and surface conductivities and relative permittivity of the constituents and depend also on the geometric characteristics of the shape inclusions (i.e., if the inclusions are spheres, spheroids or 'lentil-like') and their orientation distribution. However, we can qualitatively attribute the observed HF relaxation to the polarization of the conducting grains and correlate the LF relaxation to the presence of the insulating shells.

PEDOT:PSS films are structurally and electrically heterogeneous materials. Dc conductivity measurements provide effective values of a complex system, since long-range transport along the volume of the specimen integrates contributions of different modes of electric charge flow. Broadband dielectric measurements have the advantage of exploring the dynamics of charge carriers at different time scales and trace responses from phases of different scales.

We will show that the characteristic relaxation frequency of the HF relaxation has the same temperature variation as that of the dc conductivity *in the interior of the conducting phase*. Thus, an insight into the interiors of confined regions can be achieved, without attaching nano-probes to them. PEDOT:PSS is modeled by a suspension of conducting spheres (with conductivity $\sigma_2$, and relative static permittivity $\varepsilon_2$) covered with insulating thin shells, which are embedded into a host matrix (with conductivity $\sigma_1$ and relative permittivity $\varepsilon_1$) according to Hanai's model 13]. For the case of PEDOT:PSS films, the formulation of Hanai's theory gets simplified regarding the condition that: $\sigma_2 \gg \sigma_1$ .(and, thus, $\varepsilon_1 \sigma_2 \gg \varepsilon_2 \sigma_1$). Therefore, for any volume fraction $\Phi$, the maximum frequency $f_{max}$ *of the high frequency (HF) relaxation* and the intensity of the corresponding (given by relations (338) and (337) of Hanai's theory [13]), reduce to:

$$2\pi f_{max} = \frac{(1-\Phi)\sigma_2}{\varepsilon_0(1-\Phi)\varepsilon_2 + (2+\Phi)\varepsilon_1}$$

and

$$\Delta\varepsilon = \frac{9(\varepsilon_1\sigma_2)^2 \Phi(1-\Phi)}{[\varepsilon_0(1-\Phi)\varepsilon_2 + (2+\Phi)\varepsilon_1](1-\Phi)^2 \sigma_2^2}$$

respectively (provided that $\sigma_2 \gg \sigma_1$, which is valid for highly-conducting PEDOT grains of our system). The above two equations can combine to a single one:

$$\sigma_2 = 2\pi\varepsilon_0 \left[\frac{9\varepsilon_1^2 \Phi}{(1-\Phi)^2 \Delta\varepsilon}\right] f_{max}$$

Alternatively, $\sigma_2 = g(\Phi, \varepsilon_1, \Delta\varepsilon) f_{max}$, where $g \equiv 2\pi\varepsilon_0 \left[\frac{9\varepsilon_1^2}{(1-\Phi)\Delta\varepsilon}\right]$. The term $g(\Phi, \varepsilon_1, \Delta\varepsilon)$ is practically constant (the analysis of the dielectric data presented in the next sections, confirm that the latter constrain is true). We stress that the dielectric strengths $\Delta\varepsilon$ obtained from the fitting procedure on the present experimental stems from refers to surface conductivity data; the film thickness is quite small to extract the 'bulk' $\Delta\varepsilon$ values and, furthermore, the bulk' $\sigma_2$. However, the last equation suggests proportionality between the maximum frequency and the dc conductivity of the inclusions (for given $\Phi$ and constant g). Hence, differentiating both sides of eq. (1) with respect to $1/kT$ (k denotes the Boltzmann's constant), it results that the relaxation activation energy value coincides with the value of the activation energy for the *'local'* dc conductivity *inside* the inclusion. Finally, eq. (1) resembles the empirical Barton - Nakajima – Namikawa (BNN) relation [14, 15, 16], which asserts a proportionality between the (total) conductivity of the specimen and the maximum frequency of conductivity relaxation.

Our studies focused on the dielectric properties of PEDOT PSS at low temperatures (i.e., from T=86K to 150K)), where the two relaxation mechanisms could be better seen (the HF peak is well defined, while the LF mechanism is traced by proper fitting of dc conductivity component and a couple of Debye peaks. . In Figure 5, the natural logarithm of the frequency-independent ac conductance $G_{dc}$ (low frequency dc region) is plotted against reciprocal temperature. A linear fit to the data points yields the dc activation energy $E_{dc}^{act} = -(d\ln G_{dc}/d(1/kT)) = (0.017 \pm 0.002)eV$, where k denotes the Boltzmann's constant and T is the absolute temperature. The latter approach presumes that - within a narrow temperature region - the Arrhenius law (i.e., $G^{dc} = C\exp(-E_{dc}^{act}/kT)$, where C is a constant, holds); the good quality fitting justifies the selection of the Arrhenius equation to describe qualitatively the temperature dependence of the conductivity and, subsequently, estimate the activation energy. In the following, we will see that the Arrhenius law can describe adequately how $f_{max}$ varies with temperature. Nevertheless, the temperature region where relaxation peaks are prominent is relatively narrow and it is difficult to investigate whether other models different than the Arrhenius.

An Arrhenius equation $f_{max} = f_0 \exp(-E^{act}/kT)$, where $f_0 \equiv f_{max}(T \to \infty)$ is constant, is perfectly fitted to the data points of $\ln f_{max}(1/kT)$ diagram (Figure 6), yielding an activation energy $E_{HF}^{act} = -(d\ln f_{max}/d(1/kT)) = (0.015 \pm 0.002)eV$. As discussed in the previous section, the quantity $E_{HF}^{act}$, which is obtained by studying the temperature shift of the HT peak, is identical to the activation energy for dc conductivity *within the interior of the conducting grains*, provided that $\Delta\varepsilon$ does not vary significantly upon temperature (which is true, according the inset diagram of Figure 6). This value is lower than the macroscopic $E_{dc}^{act} = (0.017 \pm 0.002)eV$, which incorporates two competing constituents: a percolation network against energy barriers set by the insulating shells.

The data points for LF relaxation scatter, because, as mentioned earlier, fmax cannot be detected in the relaxation spectra, unless fitting procedures are employed. However, the slope of a straight line fitted to these points yields $E_{LF}^{act} = (0.15 \pm 0.02)eV$, which is an order of magnitude bigger than both $E_{dc}^{act}$ and $E_{HF}^{act}$. This finding is in agreement with the predictions of the bi-phase model of shelled spheres, according to which, the LF relaxation is correlated with the presence of the insulating shell; the poor conductivity of the shell implies an effective high energy barrier, which is recorded by the temperature shift of the LF peak. This is in accordance with a model of thermal degradation mechanism proposed for pEDOT:PSS [8]. Dc conductivity and XPS spectrograms showed that, when thermally treated, PEDOT oligomers and PSS chains separate. The latter, due to hydrophilic character concentrate at the surface of the grains, increasing the height of the potential barriers.

In conclusion, the complex impedance spectrum of PEDOT:PSS film in the frequency region 1 mHz to 1 MHz, consists of a strong dc conductivity component and a couple of dielectric relaxations. The latter are better seen when working at low temperatures (a few tens of degrees above the liquid nitrogen temperature). The presence of two relaxation peaks is in agreement with the predictions about the frequency domain response of a dense suspension of bi-phase conductive spheres covered by a thin layer of insulating shells. We showed that, within the latter dielectric theory of suspensions, we obtain information about the local dynamics of electric charge flow by studying the chance of the complex permittivity spectra at different (low temperature) isothermal conditions, and determine the activation energy for dc conductivity within the highly-conducting core of PEDOD:PSS grains. The presence of the insulating PSS shell is responsible for the low frequency dielectric relaxation peak and the activation energy related with the effective energy barrier set by the insulating PSS phase.  The methodology presented in the present work provides information about both dc conductivity along the percolation network in the (structurally and conductivity inhomogeneous) PEDOT:PSS film and local conductivity within mesoscopic phases of it.

**Figure Captions**

**Figure 1** *A vertical cross-section diagram of the cryostat (left diagram): LN: liquid nitrogen, TC: thermocouple (T-type), HPS: heating power supply and temperature control, VP: vacuum pump, S: spring,1E! and E2: leads to strip-electrodes, GND: electrical ground. On the right, a magnification of the optically flat glass plate with the attached strip-electroded is shown.*

**Figure 2** *Isotherms of the quantity $G/(\varepsilon_0 f)$ - which, as explained in the text, is proportional to $Im(\varepsilon^*)$ - vs frequency. The straight line with slope –1 fits best the room temperature (T=294K) data points.*

**Figure 3** *Fitting curve consisting of a conductivity component ($G_{dc}/f \propto f^{-1}$) (straight line) and a couple of Debye-type relaxation peaks to the data point collected at 86 K. Inset: The reduced residuals vs frequency.*

**Figure 4** *Relaxation spectra at T=86K of a PEDOT:PSS specimen heated at ambient atmosphere at 423 K for 7 h. Thermal treatment yields degradation of the dc conductivity component, permitting a clear detection of the LF dispersion (which is 'hidden' below the the strong dc conductivity component, in the more conducting virgin specimen). Inset: The reduced residuals vs frequency.*

Figure 5 *The quantity $G/(\varepsilon_0 f)$ determined from the low frequency data, whereas G is frequency independent and is therefore proportional to the dc conductivity of the film.*

**Figure 6** *Arrhenius plot for HF (full circles) and LF (full triangles) relaxations. Straight lines are best fits to the data points. Open circles: The relaxation strength Δε vs reciprocal temperature for the HF relaxation.*

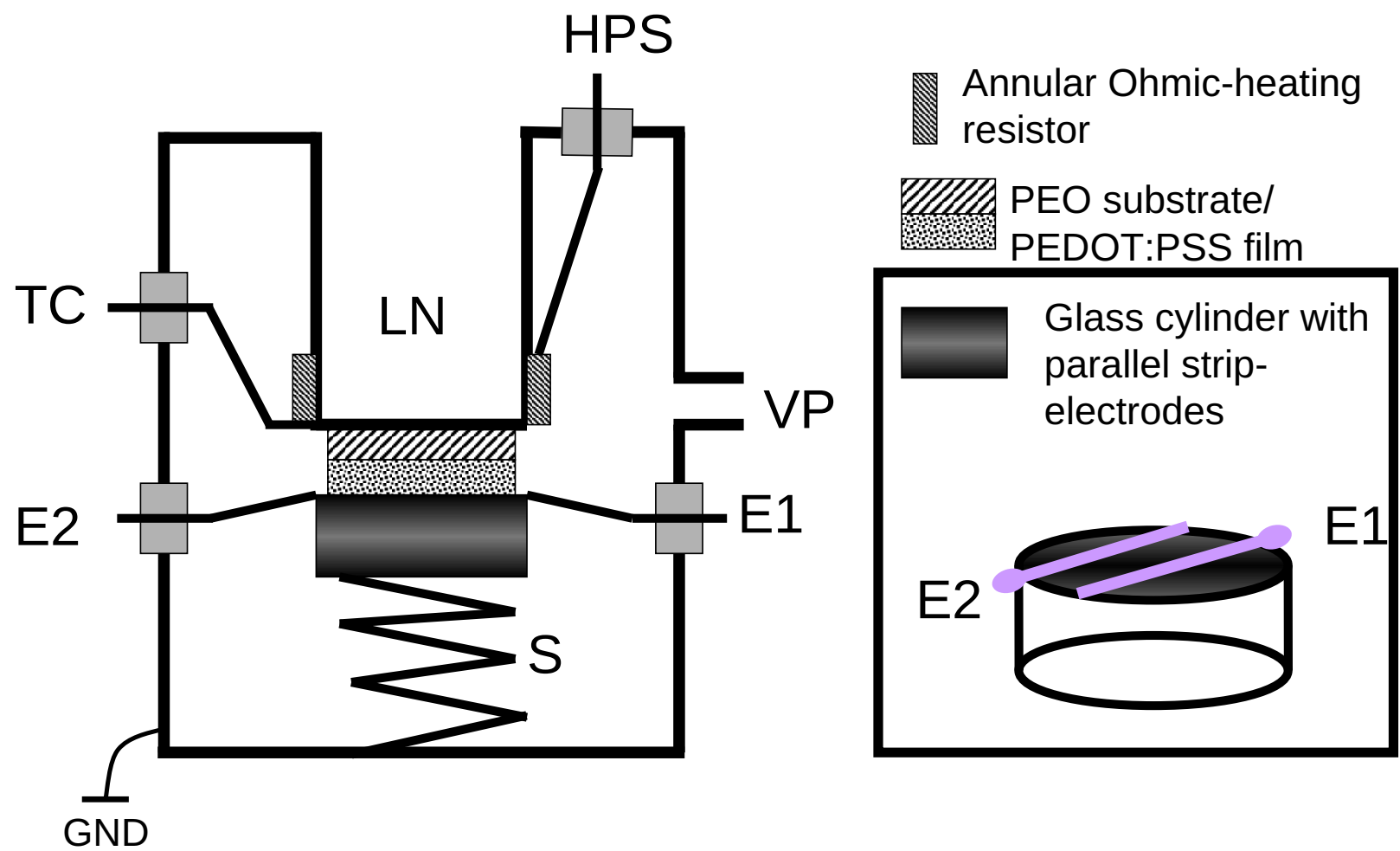

FIGURE 1

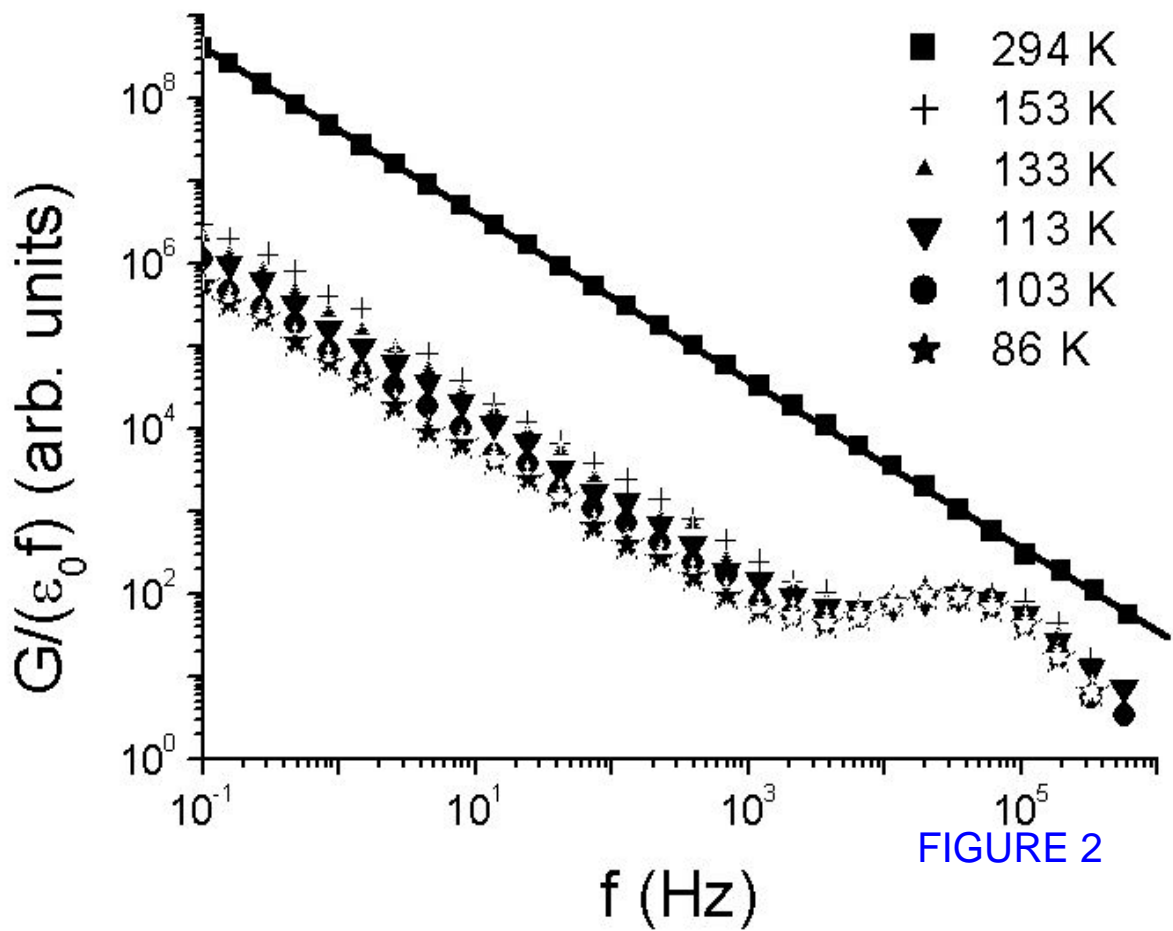

FIGURE 2

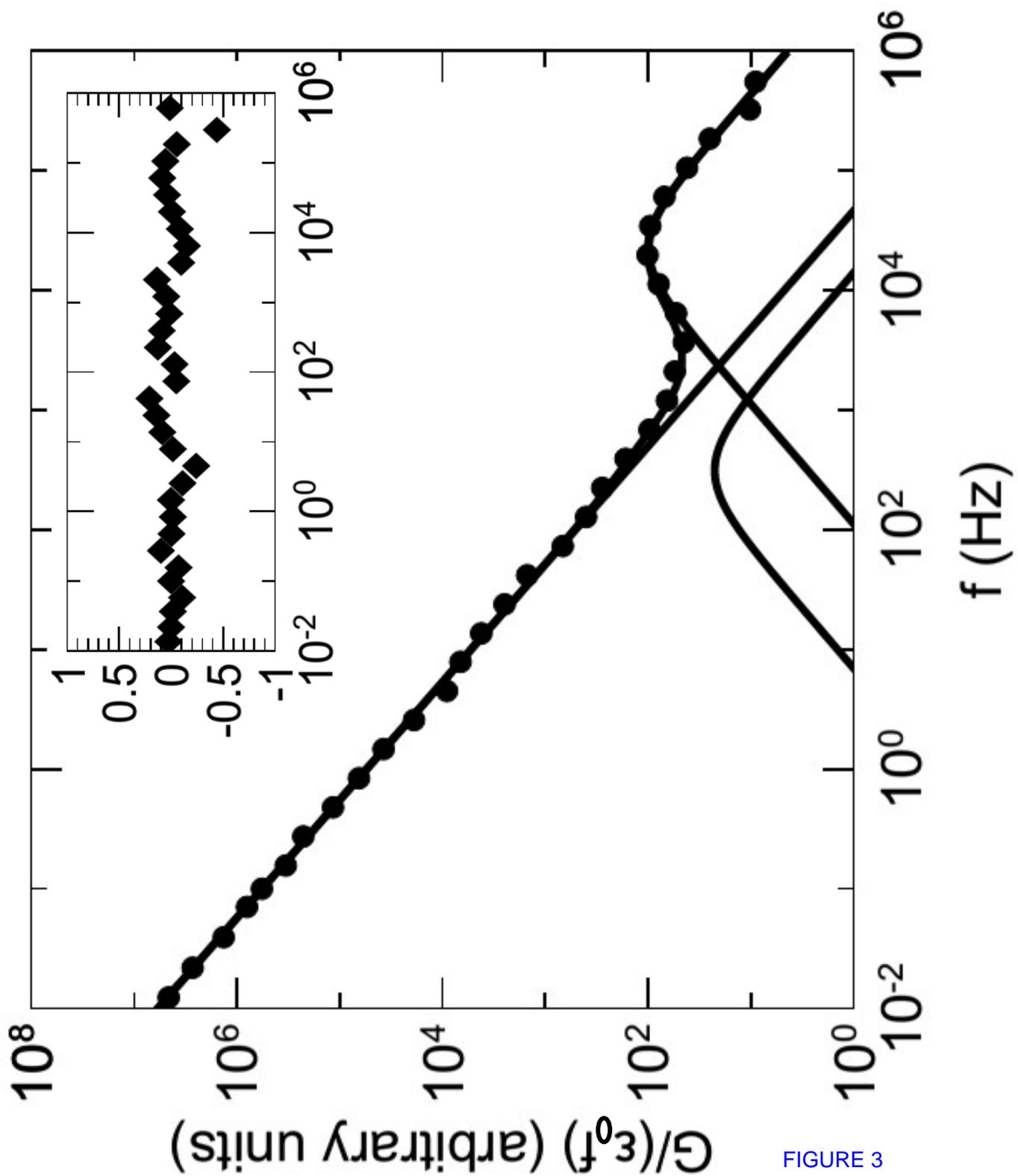

FIGURE 3

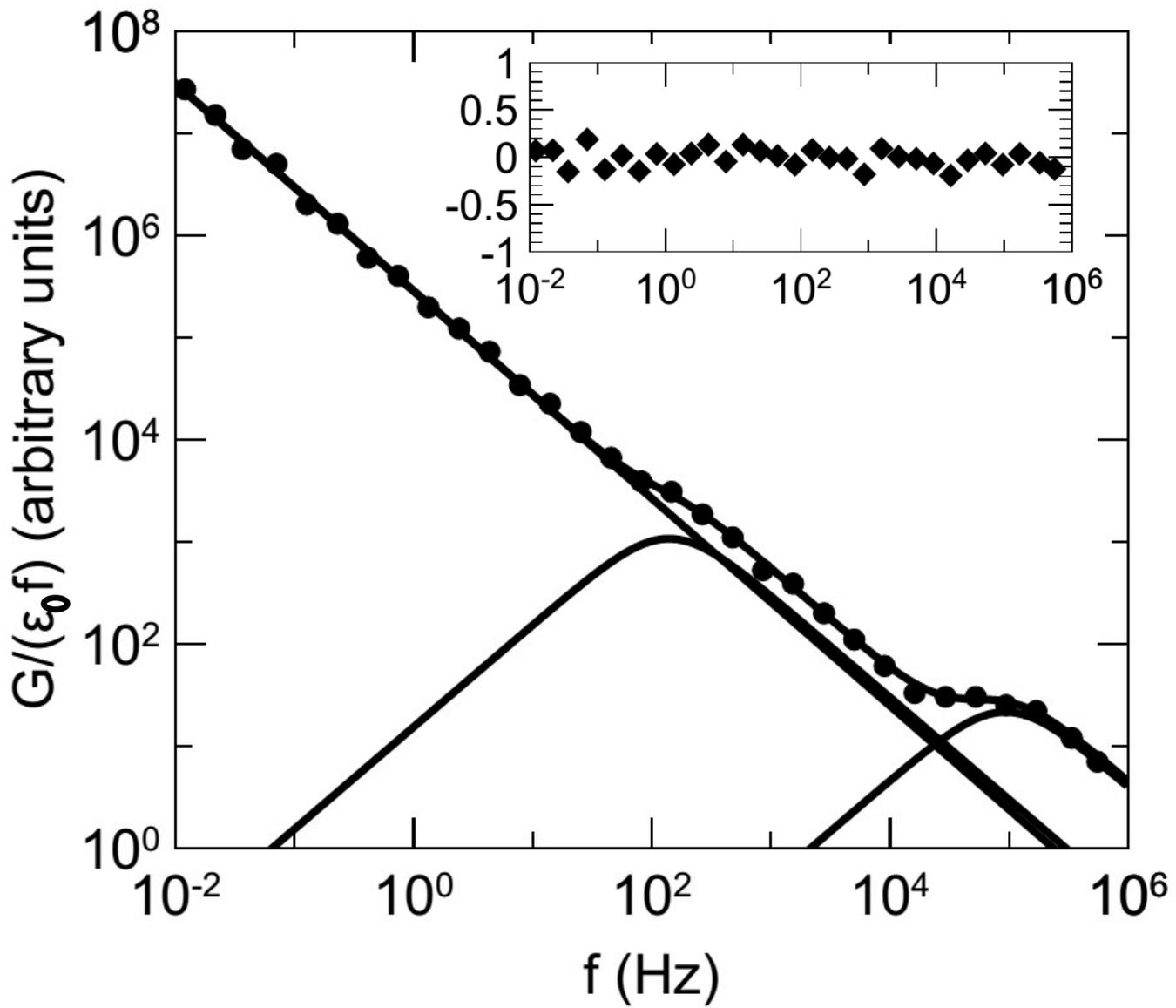

FIGURE 4

FIGURE 3

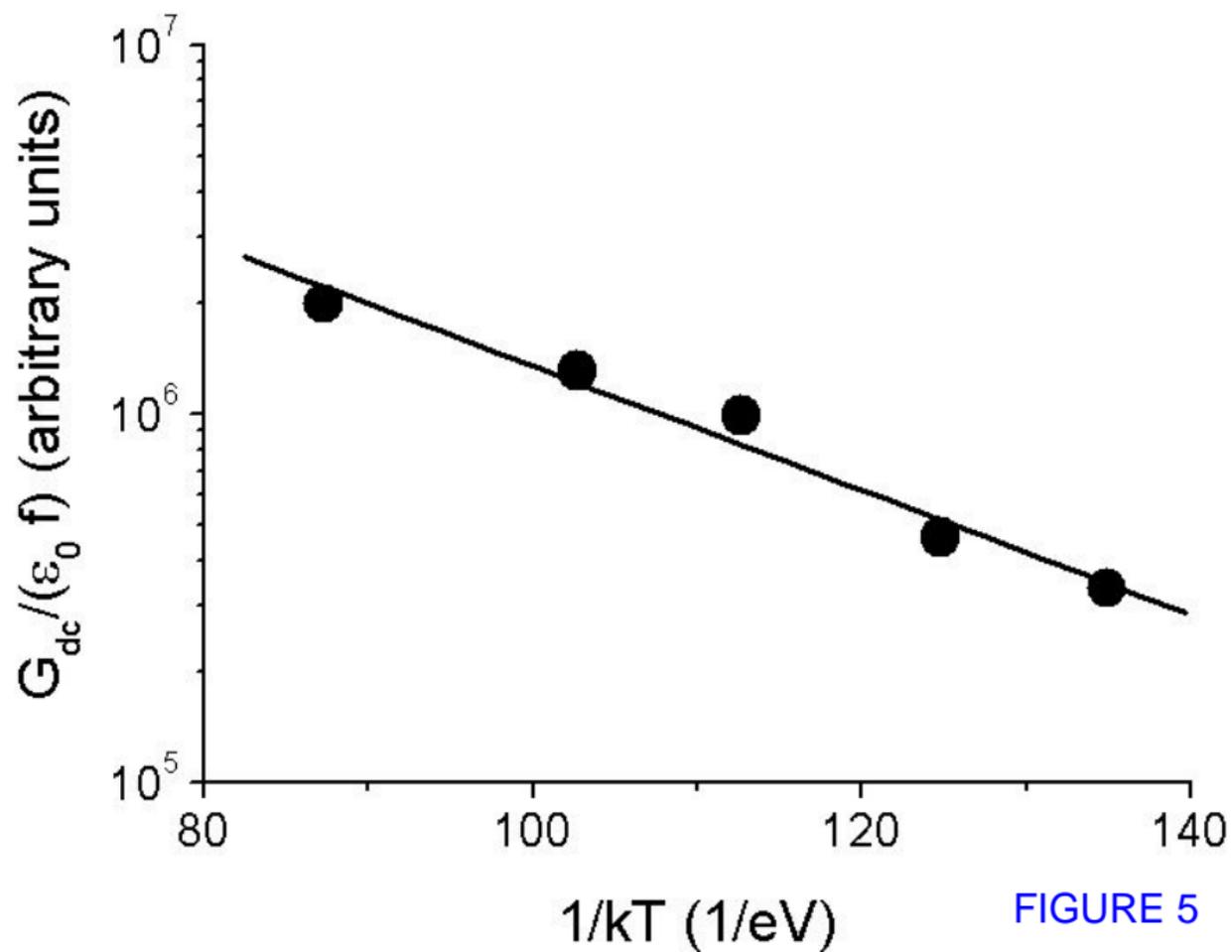

FIGURE 5

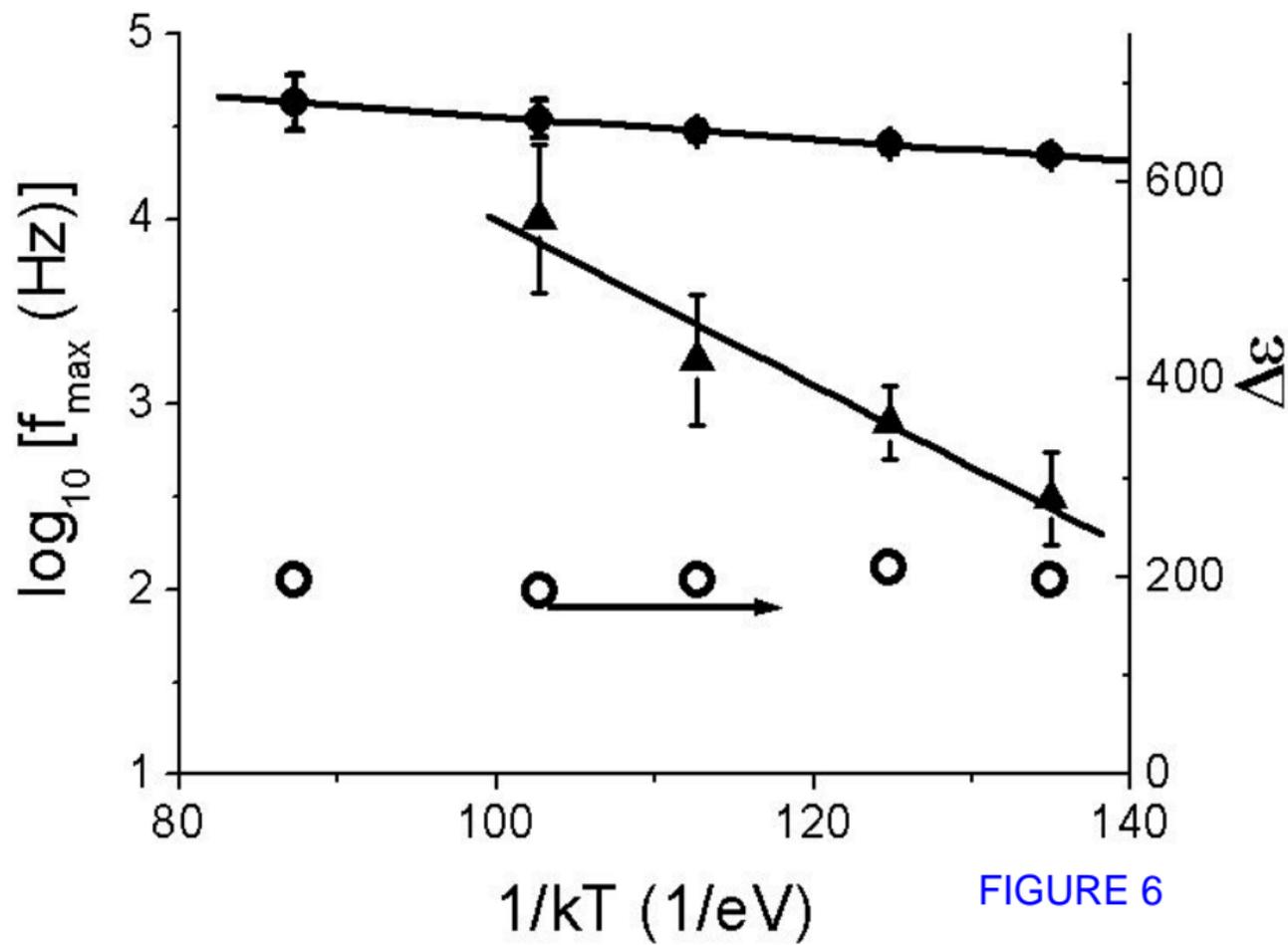

FIGURE 6